\documentclass[11pt]{article}
\usepackage[numbers, square]{natbib}
\usepackage[hang, ruled]{caption}

\usepackage{graphicx}

\usepackage{latexsym}

\def\be{\begin{equation}}
\def\ee{\end{equation}}
\def\bea{\begin{eqnarray}}
\def\eea{\end{eqnarray}}

\def\case#1/#2{\textstyle\frac{#1}{#2}}

\begin{document}

\vspace{.7in}

\begin{center}
\Large\bf {THE DEFINITION OF MACH'S PRINCIPLE}\,\footnote{It is a
pleasure to celebrate with this paper the 80th birthday of my PhD
supervisor Peter Mittelstaedt, who has taken a life-long interest in
Mach's principle.}

\vspace{.2in} \normalsize \large{\textbf{Julian
Barbour}}\,\footnote{College Farm, The Town, South Newington,
Banbury, Oxon, OX15 4JG, UK; email:
Julian.Barbour@physics.ox.ac.uk.}

\end{center}

\normalsize

\vspace{.2in}

{\textbf{Abstract.} Two definitions of Mach's principle are
proposed. Both are related to gauge theory, are universal in scope
and amount to formulations of causality that take into account the
relational nature of position, time, and size. One of them leads
directly to general relativity and may have relevance to the problem
of creating a quantum theory of gravity.

\tableofcontents

\section{Introduction}

Ernst Mach's suggestion that inertial motion is not governed by
Newton's absolute space and time but by the totality of masses in
the universe \cite{Mach1872,Mach1911,Mach1883,Mach1960} was the
primary stimulus to Einstein's creation of general relativity and
suggested its name. In 1918 Einstein \cite{Einstein1918},
\cite{Barbour1995}  (p. 185/6) coined the expression \emph{Mach's
principle} for Mach's idea and attempted a precise definition of it
in the context of general relativity. Somewhat ironically, Einstein
later disowned Mach's principle \cite{Einstein1949}, but it has
continued to fascinate researchers.

So many definitions of Mach's principle have been proposed that it
has often been dismissed as incapable of precise formulation. I
shall argue that this is because the issue at stake has not been
addressed at a sufficiently basic level. For this reason, I shall
not discuss the numerous attempted definitions but instead propose
two candidate definitions, both of which arise from very basic
considerations related to observability and the definition of
causality (more precisely, determinism) in the classical dynamics of
either particles or fields. I believe that the correct definition of
Mach's principle is important precisely because it relates to basic
issues that are likely to play an important role in the creation of
quantum gravity.

The essential content of this paper could be expressed in a quarter
of its length, but I have opted for a discursive presentation. This
is so that I can put the definition of Mach's principle in an
adequate historical perspective that distinguishes transient
concepts of `what the world is made of' from the normative
principles of empirical adequacy and causality, which, if not
eternal, have been a key part of the scientific outlook since it
came into being. The fact is that Mach formulated his principle in
terms of Newtonian masses and interaction at a distance, and
Einstein eventually came to believe that the principle was made
obsolete by the rise of field theory and local interaction. But
Einstein thereby confused ontology-independent principles with their
application to transient ontology. A key aim of this paper is to
identify first principles that do not need to be abandoned when more
superficial concepts change.

To that end, I shall begin by emphasizing that Newton introduced
absolute space and time \emph{to define velocity}, which is
displacement in unit time, as the first step in creating dynamics.
Having understood this, we can ask if there is an alternative to
Newton's absolute definition. At this point it is important to
establish what the alternative should achieve. Mach was primarily
concerned with empirical adequacy: displacement should be relative
to something observable and time must be derived from actual change.
These are ontology-independent requirements. But they are in fact
met \emph{indirectly} in Newtonian theory if properly interpreted,
as Mach came to recognize and I shall show. Mach's instinct still
told him Newtonian theory needed modification to close the gap
between its basic notions and direct empirical input, but he failed
to do this or to provide a criterion that would confirm success in
the enterprise.

This is where the formulation of causality enters the story. Drawing
on a penetrating but largely overlooked analysis by Poincar\'e
\cite{Poincar'e1902, Poincar'e1905}, I shall suggest that Mach
should have required Newtonian dynamics to be replaced by a theory
that, in a well-defined sense, has \emph{maximal predictive power}.
In other words, the theory should be as strongly causal, in a sense
that I shall make precise, as one can make it. The important thing
about this principle is that it turns out to be applicable not only
to the Newtonian ontology of masses and interactions at a distance,
but to all conceivable ontologies of the world provided only that
they are subject to continuous symmetries. The rise of the theory of
Lie groups and its widespread application in the gauge theories of
modern physics has shown that such ontologies are ubiquitous. It is
in this sense that I aim to persuade the reader of the universality
of Mach's principle as presented here. It is intimately related to
the gauge principle.

In fact, as already indicated, I shall propose two possible
definitions of Mach's principle. The first does have maximal
predictive power, while in the second that is weakened marginally. I
leave the explanation for this to the end of the paper, and merely
comment here that the alternative definition has mathematical
virtues that could outweigh its slightly weaker predictive strength.
Moreover, the alternative definition is realized in general
relativity in an intriguing manner.

\section{Newton's Argument for Absolute Space}

In 1644 Descartes published his \emph{Principles of Philosophy}
\cite{Descartes1644, Descartes1983}, in which he argued that motion
is relative. He did this in order to advance an essentially
Copernican scheme without offending the Inquisition
\cite{Barbour2001}. I will not detail here his resulting
inconsistencies but merely note that he begins by asserting among
other things that \vspace{.1in}

$\bullet$ The position and motion of any considered body is defined
only relative to other bodies.

\vspace{.1in}

$\bullet$ Since there are infinitely many bodies in the universe
that are all moving in different ways and any of these can be taken
as a reference body, any considered body has infinitely many
positions and motions.

\vspace{.1in}

He then proposed certain laws of motion, the most important of which
exactly anticipated Newton's first law: a free body will either
remain at rest or move rectilinearly at a uniform speed. He did not
attempt to reconcile this with his `official' relationalism.

At some unknown date before the mid 1680s, Newton wrote a Latin text
\emph{De gravitatione}, first published in 1962 \cite{Newton1962a}.
Much more clearly than the Scholium to the \emph{Principia} (1687)
\cite{Newton1962}, it spells out Newton's reason for introducing
absolute space (see \cite{Barbour2001}).

Like Descartes, Newton accepted that bodies move relative to each
other in space. The bodies are visible, but space is not. Newton was
clearly inspired by Descartes's idea of doing for motion what Euclid
had done for geometry: formulate axioms. More clearly than many
modern authors, to say nothing of his contemporaries, Newton saw
that the first task in such an undertaking is \emph{the definition
of velocity}. He noted especially a fatal flaw in Descartes's
relationalism: it would be impossible to say that any given body
moves rectilinearly -- the bodies used to define its motion could be
moving arbitrarily and, moreover, chosen arbitrarily.

He saw the need for the definition of what may be called
\emph{equilocality}: the identification of points that can be said
\emph{to have the same place at different times}. In a solid, each
of its points can be assumed distinguishable and fixed. Rectilinear
motion relative to them is well defined. But in the Cartesian
universe the bodies, even if assumed distinguishable, merely have
observable mutual separations $r_{ij}$ in otherwise invisible space.
One can say how the $r_{ij}$ change but not how any body moves.

Newton understood this very clearly. It is possible that he
considered the possibility of defining the motion of any particular
body relative to the totality of bodies in the universe, but he must
have dismissed that for two reasons. First, according to the
prevailing mechanical philosophy the universe should be infinite and
contain infinitely many bodies. One could never include all the
reference bodies needed to define displacement. Second, in any one
instant a given body would have a definite position relative to all
the other bodies in the universe. If they were all to remain fixed
relative to each other, the displacement of the considered body
relative to them would be uniquely defined. But, as Newton noted,
velocity is defined as the ratio of a displacement in a given time.
During this time, the reference bodies would be moving in all sorts
of different ways, making the definition of a unique displacement
virtually impossible. Faced with these seemingly insuperable
difficulties, Newton concluded his discussion of motion in \emph{De
gravitatione} by stating

\begin{quote}

\small

it is necessary that the definition of places, and hence of ...
motion, be referred to some motionless thing such as extension alone
or space in so far as it is seen to be truly distinct from bodies.

\end{quote}

\normalsize

Newton's `motionless thing' acquired the grand title \emph{absolute
space} in the famous Scholium that immediately follows the
definitions at the start of the \emph{Principia}, to which we now
turn.

\section{The Scholium Problem and Inertial Systems}

Newton acknowledged that only relative positions and times are
observable. He argued that his \emph{invisible} absolute motions
could be deduced from the \emph{visible} relative motions and
concluded the Scholium in the \emph{Principia} by claiming that how
this is to be done ``shall be explained more at large in the
following treatise. For to this end it was that I composed it.'' In
fact, he never returned to the issue in the body of the
\emph{Principia} -- and it has been remarkably neglected ever since.
Let me formulate this problem, of which Newton was acutely aware, in
modern terms:

\begin{quote}\small

\emph{The Scholium Problem.} Given only the successive separations
$r_{ij}$ of a system of particles that form a closed dynamical
system in Euclidean space and told that there does exist an inertial
frame of reference in which the particles obey Newton's laws and are
interacting in accordance with his law of universal gravitation, how
can one confirm this fact and find the motions in, for definiteness,
the system's centre-of-mass inertial system? For simplicity and
without loss of insight, it may be assumed that the particle masses
are given.

\end{quote}\normalsize

This problem, whose solution will be discussed below, could only be
neglected because nature provided \emph{material} substitutes of
absolute space and time with respect to which Newton's laws were
found to hold with remarkable accuracy: the fixed stars as a spatial
frame of reference and the rotation of the earth with respect to the
stars as a measure of time \cite{Barbour2009}. It was only in the
second half of the 19th century that scientists began to take a more
critical attitude to the foundations of dynamics and consider how
the Scholium Problem could be solved.

In fact, it was Mach's qualitative critique of Newton's concepts --
I shall come to it shortly -- that stimulated Lange \cite{Lange1885}
to attack the problem. He supposed three particles ejected from some
common point that then move freely (force-free particles) and took
their successive positions to define a material spatiotemporal frame
of reference that he called an \emph{inertial system}. By Galilean
relativity, such a system can only be determined up to Galilean
transformations. However, with respect to any such system one can
verify whether other bodies, including ones with nontrivial
interactions, do move according to Newton's laws.

Even though the notion of a force-free particle is not
unproblematic, Lange deserves great credit for this partial solution
to the Scholium Problem, and his term \emph{inertial system}, or
\emph{inertial frame of reference}, has become standard. However,
his actual method is rather cumbersome and mechanical, and a
conceptually much cleaner and more illuminating procedure had
already been proposed by Tait in 1883 \cite{Tait1883--1884}. I shall
discuss this in a generalization that takes into account not only
the relativity of position and time but also scale. The important
thing is to establish \emph{the amount of information} needed to
determine an inertial system. Most textbooks nowadays define one
simply as a frame of reference in which Newton's laws hold; they
seldom describe its actual determination or the observational input
that it needs. We shall see that this last is the key to the
definition of Mach's principle in either of the two forms that I
propose.

In the spirit of Tait's note, suppose a system of $N$ point
particles that are said to be moving inertially. We are handed
`snapshots' of them taken at certain unspecified instants by a
`God-observer'. Since only dimensionless quantities have physical
meaning, we take them to give us dimensionless separations $\hat
r_{ij}$:

\be \hat r_{ij} = {r_{ij}\over r},
~~r=\sqrt{\sum_{i<j}r_{ij}^2},\label{1} \ee
where $r_{ij}$ are the
separations measured with some arbitrary scale.

In a Cartesian representation, the particles have $3N$ coordinates,
but the $\hat r_{ij}$ contain only $3N-7$ objective (observable)
data: three are lost because the position of the centre of mass is
unknown, three because the orientation is unknown, and one because
the scale is unknown. Equivalently we can say that the positions are
known only up to Euclidean translations (3), rotations (3) and
dilatations (1). We have gauge redundancy corresponding to the
similarity group of Euclidean space.

\emph{Tait's problem} is: how can one use the snapshots to confirm
that the particles are moving inertially, and \emph{how many
snapshots are needed}? The solution to the problem is simple but
instructive.

By Galilean relativity, we can certainly take particle 1 to be
perpetually at rest at the origin of the frame that is to be found.
Next, at some instant, particle 2 will, in that frame, pass through
its least separation from particle 1, which can be taken to be the
distance 1 (this fixes the unit of distance). Further, we can choose
the coordinate axes and the unit of time such that particle 2 moves
with unit velocity in the $xy$ plane along the line $x=1, y=t$. The
coordinates of these two particles are therefore fixed to be
$(0,\,0,\,0)$ and $(1,\,t,\,0)$; we have eliminated irrelevant
ambiguity. However, at the instant $t=0$ all the other particles can
have arbitrary positions and velocities, so that a generic inertial
solution needs $6\times(N-2)=6N-12$ data to be fully specified.

We have noted that each snapshot contains $3N-7$ independent
objective data, i.e., dimensionless separations. However, the times
at which they are taken are unknown, so in fact we have only $3N-8$
real data. Thus, two snapshots contain $6N-16$ usable data, which is
4 short of the number $6N-12$ needed to construct the frame.
However, provided $N$ is large enough, three snapshots contain
enough information to construct the data and provide independent
checks that the particles are all moving inertially in a common
spatiotemporal frame.\,\footnote{Fully relational dynamics -- with
no absolute space (i.e., no a priori equilocality relation), time or
scale -- has no content unless $N\ge 3$; for $N=3$ and $N=4$ more
than three snapshots are needed to solve Tait's problem.} Note that
the determination of time is inseparably tied to the determination
of the spatial frame and that both can only be found because a
definite law of motion is operative.

Tait's result is characteristic, and it shows that for the simplest
(inertial) Newtonian problem `two-snapshot' relational initial data
do not suffice to predict the evolution. There is a four-parameter
\emph{shortfall} in the relational data. In the realistic case in
which interactions are present (gravity is never absent), there is a
five-parameter shortfall. From the Newtonian perspective, the
information that is lacking is: the value of the angular momentum
$\textbf{L}$, the amount of kinetic energy in overall expansion or
contraction of the system (because there is no absolute scale), and
the value of the dimensionless instantaneous ratio $T/V$, where $T$
is the total kinetic energy of the system and $V$ is its potential
energy.

It is a reflection on the way in which dynamics is taught that, in
my experience, even distinguished scientists struggle to get to
grips with Tait's problem and identify the reasons for the
shortfall. That it is an issue is even new to them. Note also that
4/5 of the shortfall already arises for pure inertial motion; only
1/5 arises from interactions and accelerations. This shows that,
contrary to what is frequently said, the problem of absolute vs
relative motion is not the difference between inertial and
accelerated motion. Moreover, Newton introduced absolute space to
define inertial motion, as is clear from his discussion in \emph{De
gravitatione} (see my discussion in \cite{Barbour2001}).

To conclude this section, it needs to be emphasized that there is no
epistemological defect in Newtonian mechanics; it is possible to
construct inertial systems from observed relative motions
\cite{Mittelstaedt1995}, p. 44. If there is a defect, it resides in
the curious shortfall just established, as we shall see when we come
to Poincar\'e's critique.

\section{Mach's Intuitive Critique}

Mach's contribution \cite{Mach1872,Mach1911,Mach1883,Mach1960} to
the debate about the nature of motion was threefold but mostly took
the form of suggestive comments rather than precise prescriptions.
Mach's fundamental objection was to Newton's reliance on structure
not directly derived from observation: in my terminology,
equilocality at different times and the metric of time. Although
aware of Riemann's revolutionary ideas about geometry, he did not
object to the use of Euclidean space at a given time; he frequently
talks about the observable separations between bodies in a manner
which clearly implies that these separations are compatible with
Euclidean geometry.

In fact, Mach's contribution was threefold:

\vspace{.1in}

$\bullet$ The dynamics of the universe should be described directly
and solely in terms of the changes in the observable separations.
Speaking about the Copernican revolution, he said \cite{Mach1960},
p. 284: ``The universe is not \emph{twice} given, with an earth in
rest and an earth in motion; but only \emph{once}, with its
\emph{relative} motions, alone determinable.''

\vspace{.1in}

$\bullet$ The measure of time must be derived from change: ``It is
utterly beyond our power to measure the changes of things by time.
Quite the contrary, time is an abstraction at which we arrive by
means of the changes of things'' \cite{Mach1960}, p. 273.

\vspace{.1in}

$\bullet$ The specific behaviour identified by Newton as inertial
motion could arise from some causal action of all the masses of the
universe; this could lead to some observable effects different from
Newtonian theory: ``Newton's experiment with the rotating vessel of
water simply informs us that the relative rotation of the bucket
with respect to the sides of the vessel produces no noticeable
centrifugal forces ... No one is competent to say how the experiment
would turn out if the sides of the vessel increased in thickness and
mass till they were ultimately several leagues thick''
\cite{Mach1960}, p. 284.

\vspace{.1in}

The first two of these had been anticipated in Newton's time by
Leibniz \cite{Alexander1956} and others. The third was
revolutionary. Combined with the first two, it suggests a new
concept of motion. The Newtonian view is captured in the subtitle
``How solitary bodies are moved'' of another unpublished paper `The
laws of motion' \cite{Herivel1966}, p. 208, that seems to be earlier
than \emph{De gravitatione}. By seeking laws governing individual
(solitary) bodies, Newton had no option but to assume that they move
in space as time passes. As we shall see, implementation of Mach's
principle requires us to consider, not the motion of bodies in space
and time, but \emph{positions of the universe in its configuration
space.} This claim needs some amplification.

The development of variational mechanics (described beautifully by
Lanczos \cite{Lanczos1949}) led to the notion of the configuration
space Q of a closed dynamical system. In this formalism, the points
in Q that represent the realized instantaneous configurations of the
system lie on a curve in Q called the \emph{dynamical orbit} of the
system. It should be noted that when Newtonian mechanics is
represented in this manner, the definition of Q is based on
positions in an inertial frame of reference. As we have seen, this
is problematic since such frames are obtained through a nontrivial
process. The first step in the definition of Mach's principle will
be the identification of an appropriate configuration space in which
it is to be implemented.

It will also be necessary to think about time. In the standard
accounts of dynamics, the point in Q that represents the
instantaneous configuration of the system moves along the orbit as
the time $t$ passes. But what is this $t$? If textbooks address this
question at all, it is generally said that $t$ is provided by a
clock external to the system. But if we have no access to anything
outside the considered system, as for example a solar system
surrounded by opaque dust clouds or, more relevantly, the universe,
a measure of time must somehow be extracted from within the system
itself. However, all the changes to which Mach referred are encoded
in the curve in Q. We seem to be confronted with a vicious circle --
we cannot determine the evolution parameter, which must now be
extracted from differences along the curve, before the curve itself
has been determined. Two different ways to resolve this problem will
be presented later, both involving information encoded in a curve in
Q. For the moment, I simply want to point out that Mach criticized
Newton's concept of absolute time as forcefully as the concept of
absolute space. Few people have noted this, though Mittelstaedt
coined the expression \emph{second Mach's principle}
\cite{Mittelstaedt1976, Barbour1981} in order to draw attention to
the issue.

Finally, a problem with Mach's writings -- and the cause of much
confusion and dispute -- is that he did not provide a criterion that
would establish when a theory is Machian. In fact, it is possible to
rewrite the content of Newton's laws directly in terms of the
observable separations -- I shall discuss this directly -- and thus
seemingly meet Mach's requirement without changing the physical
content of Newtonian theory. This is in fact what Lange, stimulated
by Mach's critique, showed: Newtonian dynamics can be put on a sound
epistemological basis, i.e., related to directly observable
quantities. (This, of course, presupposes that Newtonian theory is
physically correct. I shall discuss relativity later.)

Much more could be said about Mach and the way in which he has been
misunderstood, but it will be more helpful to move on directly to
Poincar\'e.

\section{Poincar\'e's Strengthened Relativity Principle}

In his \emph{Science and Hypothesis} \cite{Poincar'e1902,
Poincar'e1905}, Poincar\'e has some pertinent things to say about
the problems of absolute and relative motion, which he comments have
been much discussed in recent times. He says that it is repugnant to
the philosopher to imagine that the universe can rotate in an
invisible space but then poses the decisive question: \emph{what
precise defect}, if any, arises within Newtonian dynamics from its
use of absolute space? It is the answer to this question that will
provide us with both the definitions of Mach's principle to be
proposed in this paper.

For the purposes of his discussion, Poincar\'e assumes that a
definition of time is given and that the distances between bodies in
(Euclidean) space can be directly measured. Thus, he presupposes the
existence of a standard clock and rod. In an extension of his way of
thinking, I shall, when formulating Mach's principle, dispense with
these as in my discussion of Tait's problem. However, I shall stick
with Poincar\'e's assumptions for the moment. What he in effect said
was this. Let us suppose particles of known masses $m_i,\,i=1, 2,
..., N,$ in space between which observers can, at any instant,
observe the inter-particle separations $r_{ij}$ and, a clock being
granted, the rates of change $\dot{r}_{ij}$ of the $r_{ij}$. Thus,
at any instant the observers have access to $r_{ij},\,\dot{r}_{ij}$.
Poincar\'e also assumes that they know that the bodies satisfy
Newton's laws of motion and are governed by the law of universal
gravitation. He then asks: given such initial data, is it possible
\emph{to predict uniquely} the evolution of the $r_{ij}$, i.e., the
observable evolution?

The answer is no, and the reason is evident from the discussion of
Tait's problem: the data $r_{ij},\,\dot{r}_{ij}$ contain no
information about the angular momentum $\textbf{L}$ in the system.
That this is so is readily seen after a moment's reflection on the
two-body Kepler problem. At perihelion or aphelion any planet is
moving at right angles to the line joining it to the sun, so that
the planet--sun separation $r$ is not changing: $\dot r=0$. But the
initial data $r,\, \dot r=0$ can lead to all possible Keplerian
motions, including both circular motion and direct fall into the
sun. It is the angular momentum, invisible in the data $r,\,\dot
r=0$, that makes the difference. Since three pieces of information
are encoded in the vector $\textbf{L}$, two in its direction and one
in its magnitude, we arrive at this important conclusion: in the
generic -- and archetypal -- problem of $N$ bodies interacting with
any central forces, specification of initial data in the form
$r_{ij},\,\dot{r}_{ij}$ will leave a three-parameter
unpredictability in the evolution. When the invisibility of scale
and time are taken into account, this becomes a five-parameter
unpredictability, as is clear from the discussion of Tait's problem.

This unpredictability arises exclusively from the difference between
specifying purely relative quantities and specification in an
inertial frame of reference, for which perfect Laplacian determinism
and causality holds. The shortfall in predictive power is the price
that has to be paid for Newton's introduction of absolute kinematic
structure that is independent of the contents and relative motions
of the objects in the universe. Thus, the case for modifying
Newtonian theory is not to improve its epistemological status but
its predictive power. It is an argument from causality, not
epistemology. Before making this idea explicit, I want to point out
that Poincar\'e devised a way to characterize the nature of theories
that is different from the one that has become universal through the
manner in which Einstein created his theories of relativity.

Einstein's approach focussed all attention on the \emph{symmetries}
of the laws of nature: under what transformation laws do they retain
the same form? In his attempt to implement Mach's ideas, Einstein
sought to make the transformation laws as general as possible. He
was initially \cite{Einstein1916} convinced that \emph{general
covariance} was a powerful \emph{physical} principle but later
\cite{Einstein1918} accepted Kretschmann's argument
\cite{Kretschmann1917} that it had only formal mathematical
significance. Since then there has been much inconclusive debate
about the meaning of relativity principles and general covariance. I
do not wish to get into it here because I believe that in his
\emph{Science and Hypothesis} Poincar\'e proposed a more fruitful
and unambiguous approach. This was to define relativity in terms of
\emph{the amount of information} needed to be specified in
coordinate-independent (gauge-invariant) form if the evolution is to
be predicted uniquely. I shall strengthen Poincar\'e's formulation
in order to be able later to define Mach's principle in the
strongest and cleanest form.

Suppose we live in a Newtonian universe and consider a dynamically
isolated $N$-body subsystem within it. There are two ways in which
we can study the subsystem. The first is in the standard Newtonian
manner in an inertial frame of reference, while the second
concentrates exclusively on what is actually observable within the
subsystem.

Now, as Poincar\'e noted, certain details of the subsystem's initial
state when specified relative to its inertial system have no effect
on its subsequent directly observable evolution. Thus, one can
imagine the initial configuration of the system rotated and
translated without this having any effect. By Galilean relativity, a
uniform velocity imparted to the subsystem also has no effect. Let
us do some counting. Three numbers specify the position of the
origin of the inertial frame relative to the subsystem, three more
specify the orientation of its axes, and three more the
translational velocity of the origin. All of these have no
observable effect within the subsystem. But whereas one can boost
the origin of the inertial frame, in a passive transformation, or
equivalently the centre of mass of the subsystem, in an active
transformation, one cannot `boost the orientation'. This is because
of the dynamical effect of angular momentum, which is not encoded in
the $r_{ij},\,\dot{r}_{ij}$ but shows up in the second and higher
derivatives $\ddot{r}_{ij},...~$ as the system evolves.

This is the nub. Poincar\'e argued that this breakdown of relational
predictability is the only `defect' in Newtonian dynamics that
arises from its use of absolute space. He said that, ``for the mind
to be fully satisfied'', a strengthened form of the relativity
principle must hold: the relational data $r_{ij},\,\dot{r}_{ij}$
should determine the evolution uniquely. However, he noted with
regret that the solar system is an effectively isolated system and
manifestly does not satisfy such a strengthened relativity
principle; nature does not work the way philosophers would like.

This is the place to mention the famous study of the Newtonian
three-body problem made by Lagrange in 1772 \cite{Lagrange1772}; it
was surely at the back of Poincar\'e's mind when formulating his
critique of Newtonian dynamics. In his study, Lagrange assumed
Newtonian gravitational dynamics to be correct but then reformulated
its equations in terms of the separations $r_{ij}$ between the three
particles. He obtained three equations that contain \emph{third}
derivatives with respect to the time and thus for their solution
need specification of $r_{ij},\dot{r}_{ij},\ddot{r}_{ij}$ as initial
data. This showed that Newtonian theory with nontrivial interactions
(and not only in the case of inertial motion as in Tait's problem)
could be perfectly well expressed in terms of directly observable
quantities.$\,$\footnote{Lagrange's work was extended to arbitrary
$N$ by Betti \cite{Betti1877} and, using powerful modern techniques,
by Albouy and Chenciner \cite{Albouy1998}.} It has no
epistemological defect, as has already been noted. What remains very
curious is that when $N$ is large only very few of the second
derivatives $\ddot{r}_{ij}$ need to be specified in addition to
$r_{ij},\dot{r}_{ij}$. It is also manifestly clear that a theory in
which only $r_{ij},\dot{r}_{ij}$ need to be specified will have
greater predictive power.

In \emph{Science and Hypothesis}, Poincar\'e does not mention Mach's
ideas about the origin of inertia. As they were well known, this
surprises me. If Poincar\'e was unaware of them, this may explain
why he admitted regretfully that nature seemed to have no respect
for his strengthened relativity principle and that we would simply
have to accept what nature tells us. But Mach made it clear that we
could only hope to gain a satisfactory understanding of dynamics by
including the \emph{whole universe} in our considerations. This
opens up the possibility that the universe evolves in accordance
with a stronger relativity principle than subsystems within it. Its
overall behaviour need not be so philosophically repugnant as
Poincar\'e believed.

To conclude this part of the discussion, Poincar\'e's analysis has
two virtues: 1) it employs coordinate-free language and is expressed
solely in terms of gauge-invariant (observable) quantities; 2) it
provides the \emph{precise criterion} that Mach failed to formulate.

\section{Configurations and Gauge Redundancy}

As preparation for the definition of Mach's principle and to
demonstrate that it has universal applicability, I want to draw
attention to a widespread phenomenon in nature, introducing it by
asking this question: what empirical content underlies Euclidean
geometry?

Rods that remain mutually congruent to a good accuracy are crucial.
Using them, we can measure angles and ratios of lengths. These
dimensionless quantities have objective physical meaning. In fact, I
suspect that all quantitative measures in science reduce ultimately
to measurement of angles and length ratios. (Of course, one can
\emph{count} apples, but I am referring to measurement of their
size.) It is certainly true that ancient astronomy right up to
Kepler's discoveries relied entirely on angle measurements,
including those used to obtain time from the observed revolution of
the stars.

We can think of Euclidean geometry in the following terms. By means
of a rod, we can measure the distances between $N, N\ge 5,$
particles at some instant. We obtain $N(N-1)/2$ positive numbers,
the inter-particle distances. They could have been arbitrary, but
empirically (for $N\ge 5$ in three dimensions) they satisfy
algebraic equations (and inequalities). These empirical relations
may be called \emph{the Euclidean rules}.

They permit remarkable data compression in the representation of
facts. In a globular cluster containing a million stars there are at
any instant $\approx 10^{12}$ distances between them. All this in
principle independent information can be encoded by $3\cdot 10^6$
Cartesian coordinates. This is a colossal reduction, but it comes
with inescapable redundancy and \emph{frame arbitrariness}. One can
pass freely between Cartesian frames by Euclidean translations and
rotations; there is a $3+3=6$-fold degeneracy. In fact, since the
choice of the rod (which defines the unit of length) is arbitrary,
dilatations must be taken into account, and there is a $3+3+1$-fold
degeneracy. Thus, for $N$ particles, there are $3N$ Cartesian
coordinates but only $3N-7$ true (frame-independent) degrees of
freedom. This `economic representation with group redundancy' will
be decisive when we come to consider dynamics.

In fact, the very possibility of having dynamics is intimately
related to another facet of geometry, its `procreative' capacity.
One single set of distances between $N$ points, defining a relative
configuration, is sufficient to establish the possibility of placing
them in a Cartesian frame with coordinates $\textbf{x}_i$. The
original empirically determined distances are then given by
$r_{ij}=|\textbf{x}_i-\textbf{x}_j|$. Then infinitely many other
relative configurations can be generated by simply specifying freely
any $N$ Cartesian vectors, which form a \emph{Cartesian
configuration}, and calculating the distances between them by the
rule just given. A whole space of relative configurations can be
generated in this way. It is, however, an inescapable fact that many
distinct Cartesian configurations give rise to one and the same
relative configuration. This is so whenever two Cartesian
configurations are congruent, so that one can be carried into exact
coincidence with the other by Euclidean translations and rotations
(and scaling if we include that). \emph{A group of motions} acts on
the configurations, leaving invariant the the interparticle
separations, which we measure more or less directly and regard as
physical. In contrast, the Cartesian coordinates are \emph{gauge
dependent}, being changed by the action of the group of motions.

In the light of the above remarks, it is helpful to distinguish
three spaces. The first is the familiar Newtonian configuration
space $Q$, which for $N$ particles has $3N$ Cartesian coordinates.
If we identify all configurations in $Q$ that can be carried into
exact congruence by the transformations of the Euclidean group, we
obtain the \emph{relative configuration space} $R$. This is not a
subspace of $Q$ but a distinct \emph{quotient space}. If, in
addition, we identify all configurations that have the same shape
(adding dilatations to the group of motions), the identified
configurations form \emph{shape space} $S$, which again is a
distinct space, not a subspace of $Q$.

If we are considering a \emph{single} configuration, the economic
bonus of using Cartesian coordinates vastly outweighs any
inconvenience in the arbitrariness. However, if distances between
$N$ objects are established when they are in two \emph{different}
relative configurations there is in principle no connection between
the Cartesian frames chosen to represent each configuration. This
gives rise to serious frame arbitrariness and the first fundamental
problem of motion \cite{Barbour2001}. The second such problem
relates to time: two such configurations contain no information
about the lapse of time between them.

The `economic representation with group redundancy' discussed above
is a characteristic feature of nature that has stimulated much
mathematics, above all the theory of Lie groups and algebras.
Further examples with different groups of motion are:

\vspace{.1in}

$\bullet$ It can be established by measurement that the magnetic
field $\textbf{B}(x_i),\, i=1,2,3,$ is always such that
$\textrm{div}~\textbf{B}=0$. One can therefore represent this
empirical fact in a mathematically convenient form by introducing a
vector potential $\textbf{A}(x_i)$ such that
\begin{equation}
\textrm{curl}~\textbf{A}=\textbf{B},\label{curl}
\end{equation}
but $\textbf{A}$ carries redundant gauge information, and
$$
\textrm{curl}\,\left(\textbf{A}+{\partial\,\phi\over\partial\,\textbf{x}}\right)\equiv\textrm{curl}~\textbf{A}=\textbf{B},
$$
so that the gauge transformation
\begin{equation}
\textbf{A}\rightarrow\textbf{A}+{\partial\,\phi\over\partial\,\textbf{x}},\label{gauge}
\end{equation}
where $\phi(x_i)$ is an arbitrary scalar function, leaves the
represented empirical data unchanged. The space of 3-vector fields
$\textbf A$ with gauge redundancy can be quotiented by the symmetry
to obtain the space of gauge-invariant magnetic fields $\textbf B$.
In contrast to the spatial gauge redundancy of the first example,
this is the simplest [SU(1), Abelian] example of an internal
symmetry. The non-Abelian gauge symmetries of Yang--Mills fields are
further examples. The group of motions that here acts on the
configurations is infinite dimensional and hence a generalization of
a Lie group of motions.

\vspace{.1in}

$\bullet$ Infinitely many distance measurements between all pairs of
points in a three-dimensional manifold $\cal M$ can in principle
reveal empirically that it carries a Riemannian
3-geometry.\,\footnote{The manifold $\cal M$ is assumed to be
compact without boundary to provide the basis for a dynamically
closed universe.} This can be represented by means of a 3-metric
$g_{ij}$. However, the six components of the symmetric $g_{ij}$
include not only information about the geometry but also about the
coordinates used to represent it. Any two metrics related by a
coordinate transformation will represent the same 3-geometry, and
the six components of $g_{ij}$ will represent three geometrical
degrees of freedom and three coordinate, or gauge, degrees of
freedom. Again we have a remarkably convenient and economic
representation but with redundancy. The group of motions in this
case is the group of three-dimensional diffeomorphisms.

The configuration space with redundancy is called Riem; it is the
space of all (suitably continuous) Riemannian 3-metrics defined on
$\cal M$. The space of gauge-invariant 3-geometries is obtained by
quotienting by 3-diffeomorphisms and is called \emph{superspace}:

$$
\textrm{Superspace}:={\textrm{Riem}\over\textrm{3-diffeomorphisms}}\,.
$$

\vspace{.1in}

$\bullet$ One could also argue that all distance measurements are
local and only establish ratios. Then one would conclude that any
two 3-metrics
$$
g_{ij}(\textbf{x}),~\varphi(\textbf{x})^4g_{ij}(\textbf{x})
$$
related by a \emph{conformal} transformation
\begin{equation}
g_{ij}(\textbf{x})\rightarrow\varphi(\textbf{x})^4g_{ij}(\textbf{x}),\label{conf}
\end{equation}
generated by the positive scalar function $\varphi$ represent the
same physical configuration.\,\footnote{The purely conventional
fourth power of $\varphi$ is chosen to simplify the transformation
behaviour of the (three-dimensional) scalar curvature $R$ under
(\ref{conf}). In four dimensions, the second power is chosen for the
same reason.} This further quotienting by three-dimensional
conformal transformations leads to the infinite-dimensional space
known as \emph{conformal superspace}. It has precisely two (local)
degrees of freedom per space point.

\vspace{.1in}

These examples show that the original property of nature that gave
rise to Mach's critique of Newtonian mechanics is ubiquitous. It is
economic representation of instantaneous configurations with
redundancy and suggests that the correctly formulated Mach's
principle should be universal in scope and apply in all cases in
which such redundancy occurs. In fact, I shall argue that whenever
this characteristic redundancy is recognized in individual
configurations, it predetermines a unique dynamical theory of such
configurations that is of gauge type.\,\footnote{It is unfortunate
that the word `gauge' is used in several different senses and often
loosely, which leads to much confusion. In my view the decisive
thing is the fact that one starts with instantaneous spatial
configurations on which a group of motions acts. It generates gauge
transformations of the configurations. A dynamical theory of gauge
type arises from this fact.} In this connection, \'O Raifeartaigh
notes \cite{'ORaifeartaigh1997}, p. 13, that ``the curl property of
the magnetic field was known quite early and led, for example, to
Stokes' theorem''. In fact, coupled with the definition of Mach's
principle given later, it could have led directly with no further
input to Maxwell's equations in vacuum \cite{Barbour1982}.

It is worth noting here that physical conceptions have changed
greatly since Mach's time. It is therefore necessary to reconsider
his critique in the light of developments and above all identify the
central issue. If we accept, as is implicit in Mach's writings, that
dynamics is to be built up on the notion of configurations of the
universe, the central problem is always essentially the one that
Newton recognized so clearly: how do you define velocity in a
relational context? Once Faraday and Maxwell had introduced the
notion of fields, this problem simply became that of defining rates
of change of fields. In all cases, the problem is the same and
arises from the fact that \emph{physically} (as opposed to gauge
related) different configurations do not come with equilocality
markings or preassigned time differences between them.

In the light of this comment, it is interesting to consider a
comment that Einstein made in 1949 \cite{Einstein1949}, p. 29:

\begin{quote}

\small Mach conjectures that in a truly rational theory inertia
would have to depend upon the interaction of the masses, precisely
as was true for Newton's other forces, a conception which for a long
time I considered as in principle the correct one. It presupposes
implicitly, however, that the basic theory should be of the general
type of Newton's mechanics: masses and their interactions as the
original concepts. The attempt at such a solution does not fit into
a consistent field theory, as will be immediately recognized.

\end{quote}\normalsize

Einstein here appears insensitive to the deep problem that Newton
identified and Mach equally clearly recognized. The problem of
defining change exists in essentially the same form whatever the
ontology of the universe. The obsolescence  of Mach's ontology did
not change the underlying problem. Mach's principle needs to be
defined accordingly.

For this, we must also consider time. The above examples are all of
instantaneous \emph{spatial} configurations on which a group of
motions acts. The great work of Lie, Klein and others in the last
decades of the 19th century led to the clear recognition that the
mathematical representation of one and the same spatial structure
can be changed by the action of a group of motions. The physical
object is unchanged by this action. Minkowski and Einstein extended
this principle dramatically when they fused time with space and, in
spacetime, made it a further dimension barely distinguishable from a
spatial one. In this new picture, it was entirely natural on
Einstein's part to seek to characterize the structure of spacetime
by laws that did not depend on arbitrary coordinate representation.

However, the concept of spacetime marked a radical departure from
the `royal highroad of dynamics'$\,$\footnote{Wheeler's coining I
believe.} laid out by Newton, Euler, Lagrange, Hamilton and Jacobi.
This had led to the key concepts of the configuration and phase
spaces, in which dynamical histories are parametrized by time. All
of this work had been done in the framework of the absolute space
and time inherited, admittedly often with unease, from Newton. A few
dynamicists were just beginning to consider the Machian issues when,
in creating special relativity, Einstein and Minkowski took physics
down a new road. The ways parted. The spacetime route led to general
relativity, while Hamilton and Jacobi's synthesis of classical
dynamics led to quantum mechanics in both of its incarnations:
matrix and wave mechanics. Quantum mechanics and relativity have
since coexisted uncomfortably, as one sees in the immense difficulty
in the creation of a quantum theory of gravity.

I believe that one source of this difficulty could be Minkowski's
virtual elimination of the difference between time and space. In his
picture one cannot begin to formulate a Machian theory in which time
is derived from change. The time dimension is there from the
beginning and exists independently of the world's happenings. Of
course, Einstein modified this picture very significantly in general
relativity but not in a way in which one can readily see whether
time is derived from change.

Therefore, in order to gain a clear notion of time, there is a good
case for formulating Mach's principle in terms of a configuration
space Q, in which, as I noted earlier, a measure of time can be
extracted from the realized curve in Q. However, Q cannot be the
standard one based on absolute space. We need one that is defined
relationally by quotienting with respect to the group of motions
corresponding to the assumed ontology. As final preparation for the
definition of Mach's principle, the following analysis of velocities
defined by frames of reference will highlight the difficulties that
arise whenever the configurations of physical systems have a gauge
redundancy associated with a group of motions.

\section{The Decomposition of Velocities}

To this end, it will be sufficient to consider the velocity
decomposition that can be made in the $N$-body problem, for which I
draw on Saari's discussion \cite{Saari2005}. Let the bodies have
masses $m^a,\,a=1,2,...,N,$ position vectors
$\textbf{x}^a_i,\,i=1,2,3,$ in an inertial frame of reference with
origin permanently at the system centre of mass, and velocities
$\textbf{v}^a_i$ at some instant. Then $\textbf{x}^a_i$ and
$\textbf{v}^a_i$ can be `packaged' as $3\times N$-dimensional
vectors $\textbf{X},\,\textbf{V}$, where $X^a_i=x^a_i$ and similarly
for $\textbf{v}^a_i$. For such vectors the symmetries of Euclidean
space define a natural inner product in configuration space:
\begin{equation}
<\textbf{a}\,,\textbf{b}>=\sum_j\,m_j\,\textbf{a}_j\cdot\textbf{b}_j, j=1,2, ... , 3N.\label{InnerP}
\end{equation}
The moment of inertia $I$ and kinetic energy $T$ of the system are
then
$$
2I=<\textbf{R},\textbf{R}>\,,\,2T=<\textbf{V},\textbf{V}>.
$$
With respect to the inner product (\ref{InnerP}) there exists a
unique decomposition of $\textbf V$ relative to $\textbf{X}$ into
three orthogonal components: $\textbf{V}_r,\,\textbf{V}_d,
\textbf{V}_s$, the parts of $\textbf V$ that correspond to a
rotation of the system as a rigid body ($\textbf{V}_r$), to a
dilatation ($\textbf{V}_d$) of the system, and to a change of its
intrinsic shape ($\textbf{V}_s$). As I will shortly present an
intuitive representation of the velocity decomposition that will
simultaneously suggest how Mach's principle is to be implemented, I
refer the reader to \cite{Saari2005} for the formal proof, which
uses explicit group transformations to determine $\textbf{V}_d$ and
$\textbf{V}_r$. The intrinsic change of shape $\textbf{V}_s$ is what
remains:
\begin{equation}
\textbf{V}_s=\textbf{V}-\textbf{V}_d-\textbf{V}_r.\label{VelDecomp}
\end{equation}

It is important that the decomposition (\ref{VelDecomp}) relies on
orthogonality with respect to (\ref{InnerP}), which can only be
established when all components of the considered vectors are known
and included. The decomposition is holistic and respects Mach's
dictum that ``nature does not begin with elements'' and that
ultimately we must take everything into account (``the All'')
\cite{Mach1960}, pp. 287/8.

The velocity decomposition is important in the $N$-body problem
because the three velocity components interact as the system
evolves. In the context of Mach's principle the decomposition is
valuable because it provides a way to combine the advantages of
representation in a definite frame of reference while identifying
changes that are independent of the frame. Thus, given any
$\textbf{X}$ in an arbitrary frame and any $\textbf{V}$ in the same
frame, one can establish uniquely the part $\textbf{V}_s$ of the
latter that represents the \emph{intrinsic change} of $\textbf{X}$.
Expressed in terms of the position and momentum vectors of the
individual particles, $\textbf{x}^a$ and $\textbf{p}^a$, the parts
$\textbf{p}^a_s$ of the momenta that express pure change of shape
must satisfy

\begin{equation}
\sum_a\textbf{x}^a\times\textbf{p}^a_s=0\,.\label{Rot}
\end{equation}
\begin{equation}
\sum_a\textbf{x}^a\cdot\textbf{p}^a_s=0\,,\label{Dil}
\end{equation}

That (\ref{Dil}) and (\ref{Rot}) must hold is readily seen. For
example, under an infinitesimal dilatation $\textbf{x}^a\rightarrow
(1+\epsilon)\textbf{x}^a$, so that the corresponding momenta will be
proportional to $\textbf{x}^a$. Then all the scalar products in
(\ref{Dil}) will all be nonvanishing and have the same sign, so that
(\ref{Dil}) cannot vanish. Similarly, under a rigid-body rotation,
(\ref{Rot}) cannot vanish.

The whole mystery of absolute and relative motion is reflected in
this velocity decomposition. Suppose we have $N$ particles of known
masses and can determine the dimensionless separations (\ref{1})
between them at any instant. From the separations at any one instant
1, we can, having chosen some scale, construct a Cartesian
representation of the system. This gives us the position vector
$\textbf{X}_1$ at that instant. It is natural to place the origin of
the frame at the centre of mass. For one single configuration, we
feel no unease about choosing the orientation of the axes relative
to the configuration in any arbitrary way.\,\footnote{We could
consider choosing the axes along the principal axes of inertia but,
except in the case of rigid bodies, such a choice has no dynamical
significance and, moreover, leads to problems if the configuration
happens to be symmetric.} However, three real difficulties arise if
we now consider a second relative configuration 2 that differs
slightly from the first and wish to regard it as having arisen from
the first through motion.\,\footnote{One could equally well suppose
that the first had arisen from the second.} How is the second
configuration to be placed relative to the first? What scale is to
be chosen for it? How much time do we believe has elapsed between
the two configurations? There is nothing intrinsic in the \emph{two}
configurations that provides an answer to these questions. Different
answers lead to different velocity vectors and, in the $N$-body
problem, very different evolutions.

Let us look at this difficulty a little closer and agree to adopt
the $\textbf{X}_1$ frame and an external measure of time. We can
then let the particles move with arbitrary velocities in that frame.
But if we were to take the dimensionless relative configuration when
it is slightly different from $\textbf{X}_1$ and give it to
different mathematicians, they would be incapable of reconstructing
velocity vectors $\textbf{V}$ guaranteed to be the same. As we allow
knowledge of the masses, all they could be sure of agreeing on is
the position of the centre of mass. The orientation and scale of the
second configuration relative to the first and the separation in
time between them are not fixed.

However, there is something on which they could agree, which is the
amount, assumed infinitesimal, by which the \emph{shape} of the
second configuration has changed compared with the first. They could
do this simply by noting the changes to the dimensionless
separations (\ref{1}), but this would not suggest anything that one
could call a natural quantitative measure of the change.
Fortunately, the velocity decomposition theorem does suggest a
family of natural measures.

Represent once and for all the first configuration by a vector
$\textbf{X}_1$ in some Cartesian frame of reference. Represent the
second configuration in the same frame with its centre of mass in
any arbitrary position, with any orientation and with any scale.
This will give the vector $\textbf{X}_2$. Choose an arbitrary `time
difference' $\delta t$ and call
$\textbf{V}=(\textbf{X}_2-\textbf{X}_1)/\delta t$ the velocity of
the system. Then for all possible positions of the centre of mass,
orientation and scale of the second configuration, calculate the
scalar product $<\textbf{V},\textbf{V}>$ (note that this brings in a
weighting with the masses) and find the necessarily unique position,
orientation and scale for which this positive-definite quantity is
minimized. It is readily seen that this `best-matching' procedure,
which reduces the `incongruence' of the two figures to a minimum,
brings the centres of mass to coincidence and ensures that for the
extremalized quantities $\textbf{p}^a=m^a\delta\textbf{x}^a/\delta
t$ the relations (\ref{Dil}) and (\ref{Rot}) hold with
$\textbf{x}^a$ the position vectors of the particles in the first
configuration.

Note that the choice made for $\delta t$ merely changes the
magnitude of the \emph{intrinsic velocity} $\textbf V_s$ that
results from the extremalization, not the fact that for it the
relations (\ref{Dil}) and (\ref{Rot}) hold. Similarly, the position
in which (\ref{Dil}) and (\ref{Rot}) hold would still be the same if
we were to multiply $<\textbf{V},\textbf{V}>$ by any function of the
configuration such as its Newtonian potential energy. Because of
this freedom, we do not obtain a unique measure of change but all
such measures are natural in that they derive from the metric
(\ref{InnerP}), which in turn derives from the fundamental Euclidean
metric suitably modified to take into account the presence of
masses. The most important thing is that the intrinsic velocity
$\textbf{V}_s$ that is obtained is orthogonal to the velocity
components generated by the gauge (symmetry) transformations
whatever `best-matching' measure is chosen. Different choices of the
measure will be significant in dynamics, but at any instant they
merely change the magnitude of the intrinsic velocities, not their
directions.

We can summarize this state of affairs by saying that the extremal
$\textbf{V}_s$ is a tangent vector on shape space. It has a
magnitude and a direction. In a space of a high number of
dimensions, the direction contains much more information than the
magnitude, which is always represented by just one number.

\section{The Definition of Mach's Principle}

It is implicit, and often explicit, in Mach's writings that motion
of any individual body is to be defined with respect to the entire
universe (\cite{Mach1960}, pp. 286/7). This makes it possible to
avoid the difficulty of Cartesian relationalism, according to which
a body has as many different motions as potential reference bodies;
a multiplicity of motions is replaced by a single averaged motion.
However, the Machian view point is only possible if the unverse is a
closed dynamical system. I shall say something about the possibility
of a truly infinite universe at the end of this paper.

If we do suppose that the universe is a closed system, we can
attempt to describe it by means of a relational configuration space
obtained by some quotienting with respect to a group of motions. One
of the open issues in the implementation of Mach's principle is the
extent to which this quotienting should be taken. In all the
examples listed earlier, angles (length ratios) are taken to be
fundamental. I have three arguments, none decisive it must be
admitted, for not considering quotienting that attempts to reduce
physics to something more basic than them. The first is instinctive:
angle observations do seem to be primal. The second is that if we
take angles to be fundamental, the resulting theories will be built
up on a \emph{dimensionless} basis.\footnote{Time will not occur in
the foundations of the Machian theory, and in particle mechanics the
masses can be made dimensionless by dividing the action by the total
mass of the universe without changing the observable behaviour.
Angles and length ratios are obviously dimensionless.} This is
clearly an attractive, indeed necessary principle. My third reason
for taking angles to be fundamental is empirical: the existence of
fermions, which are described mathematically by spinors with
finitely many components. This is only possible if they are defined
relative to an equivalence class of orthonormal frames, which
presuppose an inner product. The scale (ortho\emph{normal}) is for
convenience in any particular case; it is the \emph{orthogonality}
and associated invariance of angles defined by the inner product
that is essential.\footnote{It is possible that the currently
observed fermions described by finitely many components are merely
low energy excitations of spinorial entities with infinitely many
components. These do not need an orthogonal frame for their
definition. I am indebted to Friedrich Hehl for this observation.}
Thus, I believe that there is a case for taking angles to belong to
the irreducible geometrical bedrock of physics, but we shall see
that, as of now, angles alone do not seem to be sufficient.

We are now in a position to define Mach's principle in both the
stronger and weaker forms that I propose. Its definition consists of
two parts: a) the identification of cases in which it is to be
invoked and b) the stipulation of what it is to achieve.

\vspace{.1in}

\begin{flushleft}\textbf{The Definition of Mach's Principle:} a) The
application of Mach's principle is to be considered whenever direct
observations or theoretical considerations suggest that the physical
configuration space of a closed dynamical system is to be obtained
by group quotienting of a larger configuration space that contains
redundant information unobtainable by direct observation; b) once
the quotiented configuration space $\cal Q$ has been selected, the
dynamical theory defined on it is to be such that either 1)
specification of an initial point $q\in{\cal Q}$ together with a
\emph{direction} $d$ in $\cal Q$ at $q$ defines a unique curve in
$\cal Q$ as the evolution of the system given the directly
observable initial data $q,d$ or 2) instead of a point $q$ and
direction, a point and \emph{tangent vector} at $q$ are specified.
\end{flushleft}

\vspace{.1in}

The decisive difference between both definitions and the
corresponding Newtonian law is reduction in the amount of
information needed to specify unique evolution. If one projects
Newtonian histories down to shape space, the preferred $\cal Q$ in
this case, a five-parameter family of Newtonian solutions can all
pass through any given point $q$ of $\cal Q$ in a given direction
$d$ at that point. They are initially tangent to each other, but
then `splay out' in a five-parameter family. In contrast, in the
case of 1) there is a unique curve that passes through $q$ in the
direction $d$. In case 2), a one-parameter family of curves passes
through $q$, all being tangent to each other at $q$. Although case
2) leads to a less predictive theory than case 1), which until
recently I regarded as \emph{the} formulation of Mach's principle,
it still corresponds to more predictive power than Newtonian
mechanics and has certain virtues:

$\bullet$ Vectors are amenable to mathematical manipulation in a way
that directions are not.

$\bullet$ If momentum vectors as opposed to normalized momenta
(directions) are allowed fundamental status, the momentum space of
the system has the same number of dimensions as the configuration
space. This is important in the transformation theory of quantum
mechanics and might be significant in quantum gravity.

$\bullet$ As my collaborator \'O Murchadha and I have recently shown
\cite{Barbour2010}, general relativity in the case when space is
closed without boundary satisfies condition 2).

\section{The Implementation of Mach's Principle}

Even with the strong Poincar\'e sharpening of causality and the
relativity principle that suggested the formulation of the previous
section, there are many different possible theories that implement
the above Mach's principles for particle models. However, nearly all
of them predict an anisotropic effective mass. That this could lead
to a conflict with observation was already foreseen by the first
creators of such theories: Hofmann, Reissner and Schr\"odinger
(their papers are translated in \cite{Barbour1995}). Since then, the
extraordinarily accurate Hughes--Drever null experiments
\cite{Hughes1960, Drever1961} have completely ruled out such
theories. In order to overcome this difficulty, Bertotti and I
proposed in 1982 \cite{Barbour1982} a universal method for creating
theories that implement Mach's principle based on the notion of best
matching as described above; this avoids the mass-anisotropy
problem, can be used whenever there is `economic representation with
redundancy', exploits the fundamental mathematics of Lie groups, and
is intimately related to gauge theory.

Best matching resolves the problem of defining change in the
presence of gauge redundancy in the definition of configurations but
not the representation of time (the second Mach's principle). For
the stronger form of Mach's principle, an external time is
eliminated by defining a geodesic principle on the considered
relative configuration space. In this, the line element (action) is
taken equal to the square root of an expression quadratic in the
velocities that is subjected to best matching as described above.
Then only intrinsic differences defined without any external time
contribute to the action. Since the initial condition for a geodesic
consists of a point and a direction, the strong form of Mach's
principle is implemented. Examples of this can be found in
\cite{Barbour1982,Barbour2003}, in which it is shown how Newtonian
theory can be recovered either exactly or to a very good
approximation for island universes subject to the Machian conditions
that they have exactly vanishing values for their total angular
momentum (\ref{Rot}) and dilatational momentum (\ref{Dil}).
Best-matching principles are applied to the dynamics of geometry in
superspace and conformal superspace for the stronger form of Mach's
principle in
\cite{Barbour1982,Barbour2002,Anderson2003,Anderson2005} and in
\cite{Barbour2010} for the weaker form. The cited geometrodynamic
papers present four main results. Having summarized them, I will end
the paper by considering what value they might have for future
research.

The first thing that the papers show is that general relativity and
a completely scale-invariant theory rather similar to it can be
derived from Machian first principles. Whereas application of
Machian principles to particles in Euclidean space leads to global
conditions [the vanishing of (\ref{Rot}) and (\ref{Dil}) for the
complete universe], in the theories with dynamical geometry local
conditions are obtained. Indeed, best matching on superspace leads
to general relativity, in which the $G^{00}=0$ and $G^{0i}=0,
i=1,2,3,$ members of Einstein's field equations $G^{\mu\nu}=0,
\mu,\nu=1,2,3,4,$ arise as Machian conditions. The equation
$G^{00}=0$ is a \emph{local} expression of the fact that no external
time appears in the theory, while the equations $G^{0i}=0$ arise
from best matching and are local counterparts of the condition
(\ref{Rot}).

Second, the Machian approach reverses the historical discovery of
special and general relativity and puts the derivation of gauge
theory in an interesting perspective. One first seeks to create a
Machian theory of the evolution of Riemannian 3-geometry on a closed
manifold and finds that the simplest nontrivial such theory with
\emph{local} elimination of an external time is general relativity.
If one then attempts to couple matter fields to the dynamical
geometry, it turns out that in all the simple ways in which this can
be done the matter fields must propagate with the same limiting
speed as the geometrical perturbations: the matter fields are forced
to share the same `light cone' as the geometry. In this way we find
a Machian explanation of the local validity of special relativity.
Equally striking is the fact that the simplest theories of a single
3-vector field interacting with Machian geometry and of a set of
3-vector fields interacting with geometry and among themselves turn
out to be Maxwell and Yang--Mills fields, respectively. These
results, which show that all the well-established modern dynamical
theories can be derived from Machian principles, are obtained in
\cite{Barbour2002,Anderson2002}. It should be said that some of the
uniqueness claims made in \cite{Barbour2002} relied on unjustified
tacit assumptions that ruled out more complicated possibilities, as
is noted in the careful study of \cite{Anderson2007, Anderson2007a}.
However, with the appropriate restrictions to the simplest
possibilities, the claim just made is, I believe, warranted.

Third, whereas the results so far described for Machian dynamics on
superspace could be seen as mainly of retrospective interest, those
for conformal superspace
\cite{Anderson2003,Anderson2005,Barbour2010} could be relevant to
current research. This is because of the light they cast on the
failure of general relativity to be conformally invariant (a topic
of interest ever since Weyl identified this as a problem in 1918
\cite{Weyl1918,Weyl1997}) and because the simplest Machian theory on
conformal superspace introduces a dynamically distinguished notion
of simultaneity that is nevertheless compatible with classical
general relativity and, hence, the local validity of special
relativity. This is because the symmetry breaking in the conformal
approach arises of necessity through the best-matching variation,
which imposes not only a conformal constraint but also a dynamically
imposed gauge fixing. Since the aim in quantum theory is to quantize
only true degrees of freedom, this suggests that the gravitational
degrees of freedom are precisely the two conformal degrees of
freedom within a Riemannian three-geometry. If one insists on
regarding relativity of simultaneity as fundamental, it is not
possible to identify definite degrees of freedom in this manner. It
may also be noted that theories in which spacetime symmetry is
broken by the introduction of a distinguished notion of simultaneity
have recently attracted much attention following the publication of
Ho\v rava's paper \cite{Horava2009}. Whereas the symmetry breaking
is simply imposed in Ho\v rava's approach, it arises in the
conformal approach of necessity through a dynamically imposed gauge
fixing.

Detailed discussion of the implications of the conformal results
goes beyond the scope of this paper, but I would like to draw
attention to \cite{Barbour2010}, which shows that although general
relativity in its spacetime formulation is not conformally
covariant, as Weyl noted, it can be represented as a theory on
conformal superspace, where it appears to meet Weyl's objections and
is an example of a theory in which a point and tangent vector on the
relevant quotient space determine the evolution. Only the local
shapes of space and their rate of change, expressed by a tangent
vector, play a role. They appear as the true dynamical degrees of
freedom of the theory; the local scale factor is emergent, appearing
merely in a distinguished gauge representation of the theory. The
status of the tangent vector in the initial data is also very
interesting. It is defined naturally and purely geometrically on
conformal superspace by analogy with the manner in which intrinsic
change of shape is defined in the velocity decomposition theorem
discussed in Sec. 7. It is therefore in essence a velocity vector,
which implicitly presupposes a time variable (since a velocity
vector is by definition a displacement in unit time). As implemented
in general relativity, the implicit time is also emergent and
defined by the change of the emergent scale factor.

To conclude this paper, let me say a word about the overall
structure of the universe. As we have seen, Mach's principle is
difficult to formulate if the universe has infinite extent. If we
assume a finite (spatially closed) universe, we then find that the
geometrical evolution of such a universe in accordance with Mach's
principle leads to general relativity and that Einstein's field
equations are a direct expression of the Machian nature of the
theory. These are, however, local conditions and are perfectly
compatible with a universe of infinite extent. Thus, what seems to
be a great success of the theory threatens to undermine one of the
key assumptions used to derive the theory in the first place. I will
not pretend to have an easy response to this difficulty. As of now,
the only suggestion that I can make is that quantum gravity might
dictate its resolution. Moreover, there is something intuitively
appealing about self-contained systems since, as Einstein said of
Mach's idea, it closes ``the series of causes of mechanical
phenomena'' \cite{Einstein1922}, p. 62. As we grope for a quantum
theory of the universe, a quantum Mach's principle could be a
crucial normative principle.

\vspace{.2in}.

\textbf{Acknowledgements}. Over the years many people have
contributed to the `Machian programme' described in this paper. I am
much indebted to my collaborators Bruno Bertotti, Niall \'O
Murchadha, Brendan Foster, Edward Anderson, Bryan Kelleher, Hans
Westman, Henrique Gomes, and Sean Gryb. There have also been
innumerable discussions with other people, above all Karel Kucha\v r
and Lee Smolin, who have been a constant source of valuable
comments.

\bibliography{julian_barbour_references}
\end{document}